# Aspin Bubbles

## mechanical project for the unification of the forces of Nature

Yoël Lana-Renault

Departamento de Física Teórica. Facultad de Ciencias.

Universidad de Zaragoza. 50009 - Zaragoza, Spain

e-mail: *yoelclaude@telefonica.net*
web: *www.yoel-lana-renault.es*

This paper describes a mechanical theory for the unification of the basic forces of Nature with a single wave-particle interaction. The theory is based on the hypothesis that the ultimate components of matter are just two kind of pulsating particles. The interaction between these particles immersed in a fluid-like medium (ether) reproduces all the forces in Nature: electric, nuclear, gravitational, magnetic, atomic, van der Waals, Casimir, etc. The theory also designes the internal structure of the atom and of the fundamental particles that are currently known. Thus, a new concept of physics, capable of tackling entirely new problems, is introduced.

## 1. Introduction

Our theory presented here is compatible with existing views about the nature of matter, and demonstrates that the essential properties of particles can be described in the mechanical framework of classical physics with certain assumptions about the nature of physical space, which is traditionally called the ether.

The theory is a synthesis of ideas used by Newton, Faraday, Maxwell and Einstein. In the past, the hypothesis of the ether as a fluid was decisive in the creation of the theory of the electromagnetic field. Vortex rings were used to construct a model of the atom at a time when the existence of elementary particles was not known. These days, applying vortex models to elementary particles looked reasonable. However, the present theory introduces alternative elementary particles, called *"tons"*, with left and right rotation, for constructing models of particles and anti-particles. Interactions between the tons are used to represent a charged particle, with the sign of the charge defined by the acceleration of the internal motion of the tons.

Maxwell's equations are accepted here as the basis for the description of the kinematics of the ether. As to the dynamics of the ether, it is shown that the Newtonian mass of a particle depends on an asymmetry in the size of its tons. The ether is assumed to have a nonlinear inertial behaviour.

There is a longstanding disagreement on how to approach the explanation of physical interactions. The answer appears simple when we observe a collision of tangible bodies: they interact because they touch. The fact that atoms, from which those bodies are built, actually do not touch escapes our observations. At the earliest stages of physics there was no need for a hypothesis of a field through which bodies interact. When attention focused on the phenomenon of interaction without observable contact, it was necessary either to assume a hidden mechanism of interaction in the space between the bodies or simply to accept that interaction can occur at a distance, and that forces are all we need to know about. In a way, this is a philosophical question rather than a physical one. Historically, action at a distance was well accepted, as it was quite fruitful in its results.

There is also a tradition of attempts to penetrate deeper than observable facts permit and to theorize about the physical nature of processes. Omnipresent ether was thought to be the medium for physical interactions. As far as electrodynamics goes, this approach was used by Michael Faraday, who demonstrated, even to pragmatists, that real knowledge can be gained by hypothesizing about unobservable things.

A satisfactory model of the ether remains to be worked out. For the purposes of explaining different phenomena, different models were used: fluids, solids, gases... Overall, ether does not look attractive for use in a consistent theory, but that is only if we are thinking in a framework of substances known to us. To develop ether theory, one must be free to imagine a medium with properties that may not correspond to anything we know from classical mechanics. Our goal is to understand the nature of matter and fields, even if we need to ascribe a very unusual property to space or to a substance which fills that space. The main intention of this paper is *to reduce all physical interactions to purely mechanical interactions.*

There was another use of the concept of the ether at the end of the nineteenth century, according to which everything observed not only interacted through ether but was also made of ether. This supposition was behind William Thomson's (lord Kelvin) presentations of atoms as vortex rings. This idea was further developed by J. J. Thomson. Under this approach, physical bodies are some kind of disturbance in the substance that fills space.

After the experiment of Michelson and Morley in 1887, the concept of ether as a medium was soon abandoned. The irony is that the Michelson-Morley experiment got rid of the ether-medium hypothesis but did no harm to the universal ether hypothesis.

This paper describes one specific wave-particle interaction capable of reproducing the known forces of Nature. This interaction will lead to a new concept of the atom and other developments. As the aim of the paper is to disseminate ideas taken from a broader and more general theory, only basic concepts and consequences, without intricate mathematics, are presented.

Since the early stages of fundamental physics, the origins of electricity, gravity and nuclear forces have inhabited a terrain of rather blurred words and conceptual meanings in the scientist's mind. The phenomenology of forces was somehow sidelined: only mathematical developments had a certain importance before contrasting predictions in the lab. In consequence, intricate theories have emerged from the basics of quantum mechanics and general relativity theory to explain phenomena observed at different scales in the same universe.

The solution proposed in this paper is the conclusion of a decade's search for a renewed theory of physics, which is able to explain both concepts and developments.

## 2. Statement of the theory

Nature is composed of particles called *"tons"*. These are modeled as spherical bubbles whose membranes oscillate with anharmonic motion around a point of equilibrium. This movement can be described by the following expression for the radius of the membrane:

$$r = (r_0 + A_0 \sin[\omega t])^x \qquad (1)$$

This function, that defines the oscillating motion of the spherical membrane with angular frequency $\omega$, is the exact zero-energy solution of the following anharmonic potential (L-Renault, 2000) [1]:

$$V(r) = \frac{1}{2} M x^2 \omega^2 r^2 \left[ 1 - 2 r_0 r^{-1/x} + (r_0^2 - A_0^2) r^{-2/x} \right] \qquad (2)$$

where $M$ is the mass of the membrane.

In the above expressions, $(r_o, A_o)$ are two parameters related so that $r_o > A_o > 0$. The exponent $x$ is always positive, and fulfils $1-\varepsilon < x < 1+\varepsilon$, where $\varepsilon$ an is infinitesimal.

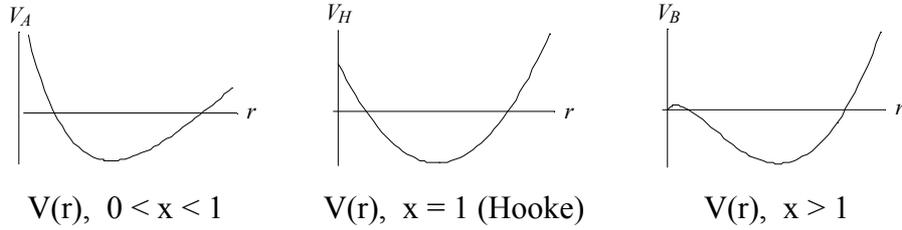

V(r), 0 < x < 1      V(r), x = 1 (Hooke)      V(r), x > 1

Figure 1

Clearly, $V(r) < E$ governs the motion of the membrane of the ton. At zero-energy, with boundary condition $V(r) = 0$, the extrema of the oscillator are:

$$R_m = r\left(-\tfrac{\pi}{2}\right) = (r_0 - A_0)^x \qquad (3)$$

$$R_M = r\left(\tfrac{\pi}{2}\right) = (r_0 + A_0)^x \qquad (4)$$

Beyond the origin, $V(r)$ has a minimum, at $r = R_1$; this is the position of equilibrium of the membrane and its value is:

$$R_1 = r(\varphi_1) = \left( r_0 + \frac{-r_0 + \sqrt{r_0^2 + 4x(x-1)A_0^2}}{2x} \right)^x \qquad (5)$$

with

$$\varphi_1 = \arcsin\left( \frac{-r_0 + \sqrt{r_0^2 + 4x(x-1)A_0^2}}{2xA_0} \right) \qquad (6)$$

The velocity of the membrane is:

$$v = \dot{r} = x \omega A_0 \cos[\omega t] (r_0 + A_0 \sin[\omega t])^{x-1} \qquad (7)$$

and its acceleration:

$$a = \ddot{r} = -x\omega^2 r\left(x - \frac{(2x-1)r_0}{r_0 + A_0 \sin[\omega t]} + \frac{(x-1)(r_0^2 - A_0^2)}{(r_0 + A_0 \sin[\omega t])^2}\right) \quad (8)$$

At the position of equilibrium ($R_1$), the speed of the membrane is maximum $v_M = v(\varphi_1)$, and $\varphi_1 \neq 0$ for $x \neq 1$. The non-zero value of $\varphi_1$ is due to the asymmetry of $V(r)$. This asymmetry is, in fact, an anharmonic correction to Hooke's potential.

Within $R_m$ and $R_1$, the membrane generates an outward force due to $a_{out} > 0$, while between $R_1$ and $R_M$ an inward force is created as a consequence of $a_{in} < 0$. Both accelerations are averaged values.

$$a_{out} = \frac{\omega}{\varphi_1 + \frac{\pi}{2}} \int_{-\frac{\pi}{2\omega}}^{\frac{\varphi_1}{\omega}} a\, dt = \frac{\omega\, v(\varphi_1)}{\varphi_1 + \frac{\pi}{2}} \quad (9)$$

$$a_{in} = \frac{\omega}{\frac{\pi}{2} - \varphi_1} \int_{\frac{\varphi_1}{\omega}}^{\frac{\pi}{2\omega}} a\, dt = \frac{\omega\, v(\varphi_1)}{\varphi_1 - \frac{\pi}{2}} \quad (10)$$

For $x = 1$, the potential is that of Hooke's Law, and $a_{out} = |a_{in}|$. Thus, the sum of both accelerations is zero and coincides with the average along one half-period. However, if $x > 1$ the total acceleration is negative, and if $0 < x < 1$, the total acceleration is positive. It takes the form:

$$a_i = a_{out} + a_{in} = \frac{2\varphi_1 \omega\, v(\varphi_1)}{\varphi_1^2 - \left(\frac{\pi}{2}\right)^2} \quad (11)$$

The above equations refer to tons that, while submerged in an isotropic fluid, produce inward or outward forces, depending on the exponent. We will call these modes *negaton B* and *positon A*, respectively. A perturbation in the density of the fluid is sufficient for the equilibrium situation to be lost and so the tons move around. Besides, the tons produce in the fluid spherical but anharmonic longitudinal waves of the form:

$$\phi(d, t) = \frac{\psi(d, t)}{d} \quad (12)$$

$$\psi(d, t) = (r_0 + A_0 \sin[kd - \omega t])^x - R_1 \quad (13)$$

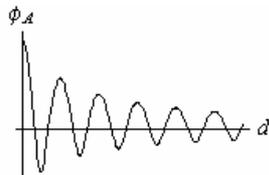
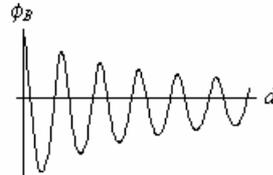

*Anharmonic wave A*   *Anharmonic wave B*

Figure 2

These waves create a polarization in the fluid with a gradient inversely proportional to the distance *d*. This would represent the electrical field. The waves **A** create a positive polarization; and the waves **B**, a negative polarization. Their speed of propagation is the velocity of light *c*.

We have found the expression for a force describing the interaction between two tons separated by a distance *d*. The nature of this force is dual, i.e. it consists of a mechanical description of the interaction and, at the same time, a physical one based on anharmonic waves. This fundamental equation is described by the following formula:

$$F_{ij}(d) = \delta_i \frac{n_j}{n_i} m_i a_j \frac{R_i R_j}{d^2 - R_j^2} \qquad (14)$$

i.e., the average force that a ton *i* exercises over a ton *j*.

The notation is as follows: $R_i$ and $R_j$ are the average radii of the membranes; $m_i$ the mass of the ton *i*; $a_j$ the total acceleration of the membrane *j*; $n_i$ and $n_j$ are fixed positive numbers; and $\delta_i$ is equal to +1 if the ton *i* is a positon or -1 if it is a negaton.

The fixed positive number *n* of a ton indicates its the quantum state. It always fulfils $n \geq 1$.

We are assuming that the force $F_{ij}$ corresponds to the self-propulsion of the ton *j* within the surrounding fluid (ether), which has a polarized gradient generated by the ton *i*. This concept of *self-propulsion* of the ton *j* is a consequence of the gradient produced by the ton *i*.

A ton has two effects on the surrounding medium called ether: a gradient and a wave. The gradient corrects the interaction forces, and the wave produces contractions and dilations of the ether that determine the polarization. This motion is anharmonic, without symmetry in the radial direction. If the ton is a positon, the total acceleration on the ether is positive and, thus, the contractions of the ether are larger than the dilations (positive polarization). If the ton is a negaton, the effect is the opposite. Thus, there is a ton *i* that generates an ether *i*, the consistency of which in the surrounding medium is determined by a gradient and a polarization. The consistency of the ether on one side of a ton *j* that is immersed in ether *i* will be different from the consistency on the other side. Consequently, the ton *j* produces different forces on each side and self-propels itself. In other words, the difference between the forces generated on each side of the ton *j* propels it.

Let us also assume the following possibility. The positon acts as a compression pump that hardens the ether. The negaton acts as a suction pump that softens the ether. In the absence of tons (at an infinite distance *d* from tons) the ether has a fixed consistency or hardness. As a positon approaches, the ether becomes harder. This hardening depends on the distance from the positon. Similarly, the effect of an approaching negaton is to soften the ether. So we can differentiate between a positon generating a more or less hard ether (but never a softer ether) and a negaton generating a more or less soft ether (but never a hard ether). See Figure 3.

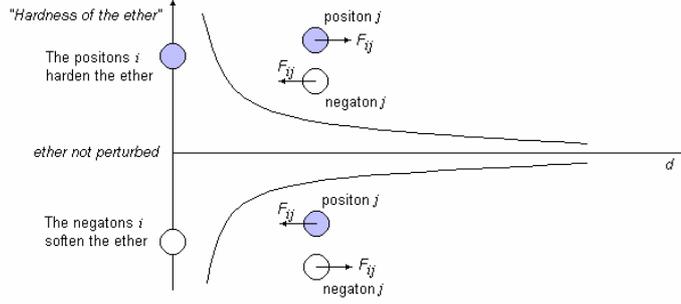

Figure 3

A possible way to deduce the expression for the force $F_{ij}$ is the following:

We do not know the structure of the ether but we can imagine that it is an elastic medium formed by a network of little cells of mass $\Delta$. The membrane of mass $M_i$ of a ton $i$ produces in this medium a gradient that we can simplify by the following dimensionless coefficient of restitution of forces. For a distance $s$ from a ton $i$ we will have (see Figure 4):

$$\chi_i(s) = \chi_0 + \delta_i \frac{M_i}{n_i \Delta} \frac{R_i}{s} \tag{15}$$

where $s \geq R_i$, $\chi_i(\infty) = \chi_o = 1/2$, and $0 \leq \chi_i(s) \leq 1$. Therefore, it is necessary that the unknown mass $\Delta$ of the ether always fulfils the condition $\Delta \geq 2M_i$.

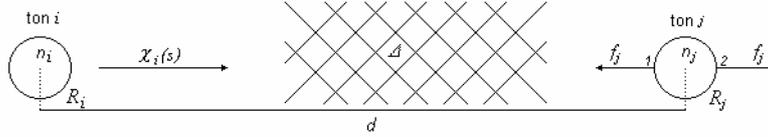

Figure 4

Besides, we can consider that the membrane of a ton $j$, that interacts with the ether $i$, always exercises over its surrounding ether the following average force:

$$f_j = (n_j \Delta) a_j \tag{16}$$

where $a_j$ is its total acceleration and $n_j \cdot \Delta$ the quantity of the surrounding perturbed ether.

Thus, applyng the principle of action-reaction to both sides (*1, 2*) in Figure 4 of the ton $j$, the average force that a ton $i$ exercises over a ton $j$ is:

$$F_{ij}(d) = \chi_i(d - R_j) f_j + \chi_i(d + R_j)(-f_j) =$$
$$= \delta_i \frac{n_j}{n_i} (2M_i) a_j \frac{R_i R_j}{d^2 - R_j^2} \tag{17}$$

We have seen that the tons are spherical bubbles whose membranes oscillate with anharmonic motion around a position of equilibrium, which produces interactions among the tons when they are immersed in the ether. Another property is that they rotate.

The angular momentum $S$ of a ton $i$ is always constant and defined by:

$$S = I_i \omega_{si} = \frac{2}{3} M_i R_i^2 \omega_{si} \tag{18}$$

where $I_i$ is the momentum of inertia of the membrane and $\omega_{si}$ the angular speed that depends on the angular frequency for the pulsing bubble, $\omega_i$.

Its magnetic momentum is:

$$\mu_i = \frac{e}{m_i} S \qquad (19)$$

By calculating the magnetic momentum of a unit of electrical charge $e$ uniformly distributed along the surface of a sphere, we obtain:

$$\mu_i = \frac{1}{3} e \, \omega_{si} \, R_i^2 \qquad (20)$$

We note that the equations (18), (19) and (20) establish a relationship between the mass of the ton and the mass of its membrane. We reach the following relationship:

$$m_i = 2 M_i \qquad (21)$$

To distinguish both masses, let us call $m_i$ active mass (Newtonian mass), the mass of the ton, and $M_i$ passive mass, the mass of the membrane. As a consequence of this, the unknown mass $\Delta$ of the ether is also a passive mass and we conclude that the average force (17) that a ton $i$ exercises over a ton $j$ is (14).

If we assume that the energy of a ton $i$ is:

$$E_i = m_i c^2 = \hbar \omega_i \qquad (22)$$

where $\hbar$ is the reduced Planck's constant, and the maximum kinetic energy of the membrane is:

$$T_i(R_1) = \frac{1}{2} M_i v_M^2 = E_i \qquad (23)$$

we obtain that $v_M$ is constant and equal to $2c$.

## 3. Resolution

The set $\{x, r_o, A_o, \omega\}$, which gives a complete description of a ton, can be calculated from its mass and number $n$. With the equation (22) we resolve the angular frequency $\omega$ for the pulsating bubble. We need three other equations.

First, we can describe the point of equilibrium of the membrane as:

$$V(R_1) + E_i = 0 \qquad (24)$$

Second, we impose a boundary condition on the interaction described in equation (14), thus forcing two tons $i$ of equal mass and value $n$ to repel each other following $F_E = ke^2/d^2$, where we assumed that $R_i \ll d$. From this condition, the equation (14) takes the following form:

$$\delta_i \, m_i \, a_i \, R_i^2 = k e^2 \qquad (25)$$

where $k$ is Coulomb's constant and $e$ the charge of the electron.

Finally, we impose that the average radius of the ton is:

$$R_i = \frac{\omega}{\pi} \int_{-\frac{\pi}{2\omega}}^{\frac{\pi}{2\omega}} r\, dt = \frac{2 n_i \hbar}{m_i c} \qquad (26)$$

With this supposition, we consider that the average radius of the ton is inversely proportional to the mass and its size depends also on the fixed positive number $n_i$ that indicates to us its quantum state. The number $n_i$ fixes the size of the ton with respect to the surrounding medium.

The equations (24), (25) and (26) can be transformed using (5), (6) and (11), so we reduce the dependencies to $\{x, r_o, A_o\}$. With these values we can fully characterize a ton.

## 4. Electric force

The equation (14) satisfies the expression for the electrical interaction between two tons, with independence of the values $n_i$, $n_j$, $m_i$, $m_j$, $R_i$, $R_j$, $a_i$ and $a_j$:

$$F_{ij}(d) = \delta_i \frac{n_j}{n_i} m_i a_j \frac{R_i R_j}{d^2 - R_j^2} = \delta_i \delta_j \frac{k e^2}{d^2 - R_j^2} \qquad (27)$$

From the above result we can describe the tons as charged particles: positive ($e^+$) for the positon, negative ($e^-$) for the negaton. In this theory the charge is not inherent to the tons, but it is a residue of the mechanical forces to which they are subjected. In other words, the theory does not assume the existence of the electrical charges. The tons are only described through the elastic properties of their membranes, and the interaction (14) fulfils the equation (25), explaining the evidence for electricity in Nature in this way. Thus, from the perspective provided by this theory and using the boundary equation (25), we can define the unitary charge $e$ by the following expression:

$$e = R_i \sqrt{\frac{\delta_i m_i a_i}{k}} \qquad (28)$$

For large distances ($d \gg R_i + R_j$) between two tons with different or same electrical polarity, the force $F_{ij}$ can be expressed as:

$$F_E(d) = \pm \frac{k e^2}{d^2} \qquad (29)$$

## 5. Gravity

It is reasonable to assume that the average radii $R_i$ for positons and negatons with same mass and number $n$ are different because the elastic properties of their membranes are different: the value of the exponent $x$ in the equation (1) is less than or greater than 1, respectively. In this section we show that a very small difference, in the asymmetry between the average radii of positons and negatons produces, for the interaction between two neutral particles composed of tons, a residual force which corresponds to the force of gravity.

Let us now define a new average radius of a ton with mass $m_i$ and value $n_i$ as:

$$R_i = \frac{2 n_i \hbar}{m_i c} Aspin_i \qquad (30)$$

where the new factor called "$Aspin_i$" is the cause of the asymmetry of the radii:

$$Aspin_i = \sqrt{1 + 2H_i + \delta_i 2\sqrt{H_i(H_i+1)}} \qquad (31)$$

with

$$H_i = \frac{G m_i^2}{k e^2} \qquad (32)$$

and $G$ being the universal constant of gravitation.

The above equation (31) implies a greater value in *Aspin* for the positon than for the negaton.

Let us assume now that a positon *A* has the same mass as the negaton *B* ($m_A = m_B$). From the equation (31), we obtain the following specific relationship:

$$Aspin_A \, Aspin_B = 1 \qquad (33)$$

For the sake of clarity, suppose a positon whose mass $m_A$ coincides with the mass $m_B$ of an electron, i.e. a positron. For this particular case, the values of their *Aspins* are:

$$Aspin_A = 1 + \mathbf{4.8989749233572340692}404886916... \cdot 10^{-22}$$

$$Aspin_B = 1 - \mathbf{4.8989749233572340692}380886961... \cdot 10^{-22}$$

These values have a negligible effect on the average radius $R_i$, but their relevance to gravity is essential.

The asymmetry of the radius allows to obtain a definite relationship between the mechanical interaction $F_{ij}$ and the electrical forces to be obtained:

$$F_{ij}(d) = \delta_i \frac{n_j}{n_i} m_i a_j \frac{R_i R_j}{d^2 - R_j^2} = \delta_i \delta_j \frac{Aspin_i}{Aspin_j} \frac{k e^2}{d^2 - R_j^2} \qquad (34)$$

As consequence, for large distances between two tons ($d \gg R_i + R_j$) with different mass and number *n*, we obtain a very small difference between the forces $F_{ij}$ and $F_{ji}$. From (34) we obtain the following relation:

$$\frac{F_{ij}}{F_{ji}} = \left(\frac{Aspin_i}{Aspin_j}\right)^2 \qquad (35)$$

## 6. ∑$F_{ij}$ as force of gravity

The values of *Aspins* translate the asymmetry of the average radii into the forces $F_{ij}$. As a consequence of this, the application of $F_{ij}$ to two neutral particles *M* and *M'* leads us to a residual force. Indeed, the total force produced by *i*-type tons with a mass *M* over a *j*-type tons with a mass *M'* corresponds to the force of gravity.

For the sake of simplicity, suppose that we create a mass *M* with one positon *A* and one negaton *B*, on the condition that $m_A + m_B = M$, and another different mass *M'* on the condition that $m_a + m_b = M'$. Using the formula (34) and neglecting the average radii $R_i$, $R_j$ in relation to the distance *d*, we obtain the force of gravity:

$$F_{MM'} = \sum F_{ij} = F_{Ab} + F_{Aa} + F_{Bb} + F_{Ba} = \tag{36}$$

$$= \frac{k e^2}{d^2} \left( -\frac{Aspin_A}{Aspin_b} + \frac{Aspin_A}{Aspin_a} + \frac{Aspin_B}{Aspin_b} - \frac{Aspin_B}{Aspin_a} \right) = -G \frac{MM'}{d^2}$$

It is also verified that $M'$ attracts $M$ with the same force:

$$F_{M'M} = \sum F_{ji} = -G \frac{M'M}{d^2} \tag{37}$$

This can also be checked for any neutral particles $M$ and $M'$ made of multiple tons.

*Caution*: due to the big differences between electrical and gravity forces, to obtain results in this theory, it is necessary to calculate always with a minimum of 70 significant digits.

## 7. Various properties of the tons

Once the new set of values $\{x, r_o, A_o, \omega\}$ are known with the new value of $R_i$, we can infer every characteristic of a ton.

To compute the total acceleration $a_i$ of its membrane, we can used the formula (11) or solved (25) using (30):

$$a_i = \frac{\delta_i m_i k e^2 c^2}{4 n_i^2 \hbar^2 Aspin_i^2} \tag{38}$$

Then the relation between accelerations of two tons is

$$\frac{a_i}{a_j} = \delta_i \delta_j \frac{n_j^2}{n_i^2} \frac{m_i}{m_j} \frac{Aspin_j^2}{Aspin_i^2} \tag{39}$$

and the relation between the average radii of two tons is:

$$\frac{R_i}{R_j} = \frac{n_i}{n_j} \frac{m_j}{m_i} \frac{Aspin_i}{Aspin_j} \tag{40}$$

Using (33), positons *A* and negatons *B* with same mass and $n$ always satisfy:

$$\frac{R_A}{R_B} = \sqrt{\frac{-a_B}{a_A}} = Aspin_A^2 \tag{41}$$

Finally, calculating the extreme radii $R_M$ (4), $R_m$ (3), the average radius $R_i$ (30) and the position of equilibrium $R_1$ (5) for positons *A* and negatons *B* with same mass and different or same value $n$, they always fulfil:

$$R_{MA} + R_{mB} \cong R_{MB} + R_{mA} \cong R_B + R_A \cong R_{1B} + R_{1A} \tag{42}$$

## 8. Nuclear force

At very short distances ($d \approx R_i + R_j$), $F_{ij}$ increases considerably due to the presence of the term $d^2 - R_j^2$ in equation (34). For two opposed tons forming a neutral composed particle, $F_{ij}$ is null because of the centrifugal force. In this situation the interaction $F_{ij}$ establishes an extremely strong link comparable to the nuclear forces. In other words, $F_{ij}$ plays the role of a nuclear force and will allow us to explain the constitution of matter.

## 9. Modeling particles and nuclei

We have established that the tons can model the elementary structure of known matter. Without external forces, free negatons are equivalent to the leptons (electron, muon, tau). Their corresponding anti-particles are the free positons. The theory incorporates the possibility of finding new leptons.

Opposed tons are linked by $F_{ij}$. Nuclei and particles present in Nature are constructed from restricted values of $n$.

### 9.1 Neutron

The neutron is composed of two opposed tons of the same mass and numbers $n_A$, $n_B$. Each ton has half the mass of the neutron, and the same internal energy and frequency. When forming any particle, positon and negaton are in orbital opposition precessing and pulsing with a phase-shift of 180º, reaching maximum stability with the minimum size. The weakness of its $F_{ij}$ is responsible for the neutron's instability. Antineutrons are obtained by exchanging the tons, as happens with the rest of antimatter.

### 9.2 Neutrino

Its configuration is similar to the neutron's, but the motions of the tons are different.

### 9.3 Proton

The proton is modeled with three tons, two positons in orbit around a negaton. The mass of each ton is one-third of the total, and they have numbers $n_A$, $n_B$. The positons precess around the negaton. Between positons and negaton there is a shift of 180º. The interaction $F_{ij}$ is strong. The theory predicts the proton as a very stable particle.

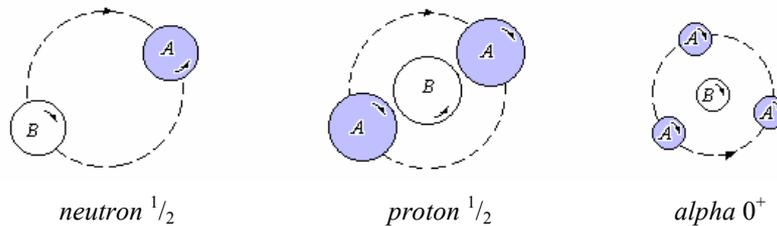

*neutron* $^1/_2$     *proton* $^1/_2$     *alpha* $0^+$

Figure 5

### 9.4 Nuclei

By adding positons and negatons, the theory can continue to model nuclei. It predicts that a nucleus with charge $z|e|$ and mass $A$ has "$2z-1$" positons linked to "$z-1$" negatons. In other words, the total number of particles is "$3z-2$" tons and the apparent charge is $z|e|$. The only exception to this model is hydrogen.

All the tons orbit in the same plane, precess and have the same internal energy and frequency. Thus, in the theory the mass is distributed evenly among the tons. As there is also a shift of 180º between positons and negatons, nuclear dimensions are minimum according to a geometrical distribution of forces, numbers $n_A$, $n_B$, angular momenta, etc., within a distribution of a sequence of spherical layers which explain the known properties of nuclei.

The construction of the different nuclei available in Nature is governed by the need to couple positons and negatons orbiting with the above-mentioned phase-shift, i.e. the variation of the dimensions of the nucleus because of their motion.

Anti-nuclei are identical to nuclei, with the only difference that the "$2z-1$" negatons are linked to "$z-1$" positons.

Isotopes have the same geometrical configuration because they have the same amount of tons. However, their masses, values $n$ and momenta change, as well as their dimensions and $F_{ij}$.

## 10. Dimensions of the tons

For the electron and positron, in their first quantum state ($n_B = 1$, $n_A = 1,001548058...$) we obtain:

$x_B = 1,0011274911...$ $\qquad$ $x_A = 0,9988785280...$

$r_{oB} = 7,9668703886... \cdot 10^{-13}$ $\qquad$ $r_{oA} = 7,4992787908... \cdot 10^{-13}$

$A_{oB} = 7,9606412203... \cdot 10^{-13}$ $\qquad$ $A_{oA} = 7,4935178639... \cdot 10^{-13}$

$R_{MB} = \mathbf{1,54}47064381... \cdot 10^{-12}$ m $\qquad$ $R_{MA} = \mathbf{1,54}57635652... \cdot 10^{-12}$ m

$R_{IB} = \mathbf{7,729}2113844... \cdot 10^{-13}$ m $\qquad$ $R_A = \mathbf{7,73}51424500... \cdot 10^{-13}$ m

$R_B = \mathbf{7,72}31865093... \cdot 10^{-13}$ m $\qquad$ $R_{IA} = \mathbf{7,729}1324811... \cdot 10^{-13}$ m

$R_{mB} = \mathbf{5,9}880568863... \cdot 10^{-16}$ m $\qquad$ $R_{mA} = \mathbf{5,9}921551196... \cdot 10^{-16}$ m

We should notice that the first quantum state ($n = 1$) for the electron is possible in the resolution of the equations (24), (25) and (30) but not for the positron. The quantum state $n_B = 1$ is a limit in the resolution of these equations for all the negatons. The positons have an other limit whose value is $1 < n_A < 1,001...$ To obtain similar dimensions to both tons, we have chosen a particular value $n_A$ such that $R_{MA}/R_{mA} \cong R_{MB}/R_{mB} = 2579,645...$

For the quantum state $n_B = n_A = 2$ we obtain:

$x_B = 1,0005630473...$ $\qquad$ $x_A = 0,9994377032...$

$r_{oB} = 1,5684014592... \cdot 10^{-12}$ $\qquad$ $r_{oA} = 1,5212381229... \cdot 10^{-12}$

$A_{oB} = 7,8378791349... \cdot 10^{-13}$ $\qquad$ $A_{oA} = 7,6101928933... \cdot 10^{-13}$

$R_{MB} = \mathbf{2,3169}938036... \cdot 10^{-12}$ m $\qquad$ $R_{MA} = \mathbf{2,3169}180863... \cdot 10^{-12}$ m

$R_{IB} = \mathbf{1,544}7984001... \cdot 10^{-12}$ m $\qquad$ $R_A = \mathbf{1,5446373018}... \cdot 10^{-12}$ m

$R_B = \mathbf{1,5446373018}... \cdot 10^{-12}$ m $\qquad$ $R_{IA} = \mathbf{1,544}4761953... \cdot 10^{-12}$ m

$R_{mB} = \mathbf{7,72}39582967... \cdot 10^{-13}$ m $\qquad$ $R_{mA} = \mathbf{7,72}24149366... \cdot 10^{-13}$ m

such that $R_{MB}/R_{mB} \cong R_{MA}/R_{mA} \cong 3$

For the negaton and positon with masses $m_B = m_A = 10^3 \cdot m_o$ ($m_o$, the mass of the electron), in their first quantum state ($n_B = 1$, $n_A = 1{,}001548058...$) we obtain:

$x_B = 1{,}0011274911...$          $x_A = 0{,}9988785280...$

$r_{oB} = 8{,}0290916587... \cdot 10^{-16}$          $r_{oA} = 7{,}4413427031... \cdot 10^{-16}$

$A_{oB} = 8{,}0228138407... \cdot 10^{-16}$          $A_{oA} = 7{,}4356262826... \cdot 10^{-16}$

$R_{MB} = \mathbf{1{,}54}47064381... \cdot 10^{-15}$ m          $R_{MA} = \mathbf{1{,}54}57635652... \cdot 10^{-15}$ m

$R_{1B} = \mathbf{7{,}729}2113844... \cdot 10^{-16}$ m          $R_A = 7{,}7351424500... \cdot 10^{-16}$ m

$R_B = 7{,}7231865093... \cdot 10^{-16}$ m          $R_{1A} = \mathbf{7{,}729}1324811... \cdot 10^{-16}$ m

$R_{mB} = \mathbf{5{,}9}880568863... \cdot 10^{-19}$ m          $R_{mA} = \mathbf{5{,}9}921551196... \cdot 10^{-19}$ m

We note that the exponents $x$ do not depend practically on the masses and besides, the values of the radii are those of the electron and positron multiplied by $10^{-3}$.

## 11. Atom, radioactivity and nuclear reactions

In this theory, radioactivity is explained as the probabilistic scattering of one neutrino against either a particle or a nucleus with weak $F_{ij}$ (unstable). Let us define a particle $X$ with atomic mass $A$ and charge $z$ as:

$$X_{z,\,negatons}^{A,\,positons}$$

Conserving the total number of tons involved in a collision, nuclear reactions can be described with the following rules:

$$neutrino_{0,\,1}^{m,\,1} + neutron_{0,\,1}^{1,\,1} = H_{1,\,1}^{1,\,2} + electron_{-1,\,1}^{m_o,\,0}$$

$$neutrino_{0,\,1}^{m,\,1} + H_{1,\,1}^{3,\,2} = He_{2,\,1}^{3,\,3} + electron_{-1,\,1}^{m_o,\,0}$$

$$antineutrino_{0,\,1}^{m,\,1} + N_{7,\,6}^{12,\,13} = C_{6,\,5}^{12,\,11} + positron_{1,\,0}^{m_o,\,1} + \left(2\,X_{0,\,1}^{m,\,1}\right)$$

$$neutrino_{0,\,1}^{m,\,1} + U_{92,\,91}^{238,\,183} = Th_{90,\,89}^{234,\,179} + He_{2,\,1}^{4,\,3} + \left(2\,X_{0,\,1}^{m,\,1}\right)$$

$$neutron_{0,\,1}^{1,\,1} + U_{92,\,91}^{235,\,183} = Xe_{54,\,53}^{140,\,107} + Sr_{38,\,37}^{94,\,75} + 2\,neutron_{0,\,1}^{1,\,1}$$

where the parentheses indicate particle and anti-particle sets.

An example of an electronic capture could be:

$$Be_{4,\,3}^{7,\,7} + electron_{-1,\,1}^{m_o,\,0} = Li_{3,\,2}^{7,\,5} + \left(2\,X_{0,\,1}^{m,\,1}\right)$$

Within an atom, the interferential field produced by the tons of the nucleus determines the positions of the electrons. In the case of hydrogen, the electron, due to its self-propulsion and its restricted values $n$, only orbits around the proton (2 *positons* + 1 *negaton*) in certain stable orbits (quantum states) in equilibrium. In other words, the electron orbits in a situation of constant equilibrium of forces, like a circular tunnel produced by the contractions and dilations of ether as consequence of the presence of the nucleus. The number $n$ fixes the size of the electron to the surrounding ether.

Positons and negatons are geometrically ordered in concentric layers forming the nucleus. The configuration of the atom includes electrons in the same plane as the nucleus, and electrons in polar positions causing precession of the atom.

## 12. Attraction between two atoms

We calculated the interaction $\Sigma F_{ij}$ between two hydrogen atoms. Because the two positons of the proton and the electron (negaton) orbit and precess around the central negaton, we considered the distance $d$ equal for all the interactions $F_{ij}$.

At very short distances, $10^{-10} \leq d \leq 10^{-4}$ m, we obtained a stronger attraction force that is inversely proportional to the distance $d$ to the power of four. The expression obtained is:

$$\sum F_{ij} = \sum F_{ji} = \frac{cons.}{d^4} \qquad (43)$$

with $cons. = -1{,}238\ldots\cdot 10^{-70}$ N m$^4$ to $n_i = n_j = 1$.

Between distances $10^{-4}$ and $10^{-3}$ m, the force decreases slowly until the force of gravity is reached. Finally, for distances always greater than one millimeter, $\Sigma F_{ij}$ is the force of gravity.

With others atoms, we obtain similars results. The explanation is the following:

If in the formula of the gravity

$$F_{MM'} = \sum F_{ij} = \sum \delta_i \, \delta_j \, \frac{Aspin_i}{Aspin_j} \, \frac{k\,e^2}{d^2 - R_j^2} \qquad (44)$$

we develop

$$\frac{1}{d^2 - R_j^2} = \frac{1}{d^2} + \frac{R_j^2}{d^4} + \frac{R_j^4}{d^6} + \ldots \qquad (45)$$

and take the two first terms, we obtain

$$F_{MM'} = \sum F_{ij} = \frac{k\,e^2}{d^2} \sum \delta_i \, \delta_j \, \frac{Aspin_i}{Aspin_j} + \frac{k\,e^2}{d^4} \sum \delta_i \, \delta_j \, R_j^2 \, \frac{Aspin_i}{Aspin_j} \qquad (46)$$

The first term is the force of gravity for all distances $d$. The second term is always negligible excepting the interactions $(i, j)$ between tons from atom $i$ and electrons from atom $j$ for distances $10^{-10} \leq d \leq 10^{-4}$ m. The reason of this resides in the average radii ($R_j$) of the electrons that are always bigger than the tons of a nucleus (see relation 40). Therefore, if we denominate the electrons from atom $j$ with the subscript $jb$, the result (43) is the general expression to the force existing between any two atoms for small distances ($10^{-10} \leq d \leq 10^{-4}$ m) where the constant has the value:

$$cons. = -\frac{k\,e^2}{Aspin_{jb}} \sum \delta_i \, R_{jb}^2 \, Aspin_i \qquad (47)$$

and if all the electrons $jb$ have the same number $n$, then, we have

$$cons. = -\frac{k\,e^2}{Aspin_{jb}} R_{jb}^2 \sum \delta_i \, Aspin_i \qquad (48)$$

All the attraction forces among atoms or molecules (ionic, covalent, van der Waals, Casimir, etc.) have their origin in the summation of the forces $F_{ij}$.

# 13. An attempt on electromagnetism. Biot-Savart Law and the Magnetic force of Lorentz

Let us suppose the ether with some viscosity, being stretched and tensed by the motion of the ton while rotating. Let us also suppose that the speed of propagation is *c*. We also make the following assumptions (a, b):

a) A positon non-linked with velocity **v** drives the ether in a clockwise sense. Its spin always has the same direction as the trajectory.

b) In the case of a negaton, it drives the ether counter-clockwise. The spin has the same direction, but in the opposite direction to the trajectory.

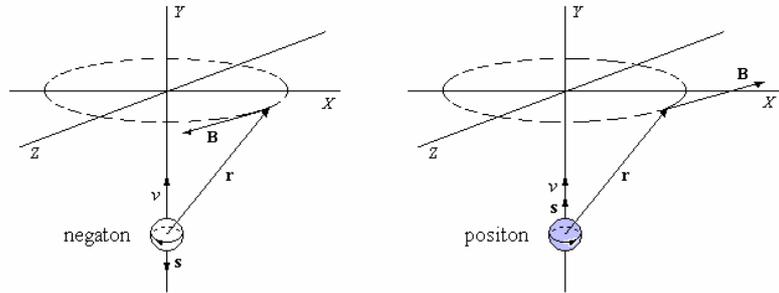

Figure 6

Under these conditions, we can generalize the law of Biot-Savart in the following way:

$$\vec{B} = \frac{\mu_0}{4\pi} \frac{ev}{\vec{r}^3} \hat{s} \wedge \vec{r} \qquad (49)$$

where the magnetic field **B** indicates the characteristics of the stretched ether by a ton at a distance **r**, velocity *v* and direction **s** of the spin.

Then, a ton with velocity *v* in the above ether **B** is subject to a force **F**:

$$\vec{F} = ev\,\hat{s} \wedge \vec{B} \qquad (50)$$

According to the following figure, we can decompose spin **S** into two components in the plane determined by **S** and **B**: $S_y$ and $S_x$. $S_y$ always stretches and tenses the ether in the perpendicular direction of **B** and is the same in all directions (equilibrium). However, $S_x$ stretches and tenses the ether more in the direction and sense of **B** (side *1* of the ton). On the other side (*2*), the effect is the opposite. Thus, this anisotropy of the ether produces the above expression of the magnetic force of Lorentz **F** (50).

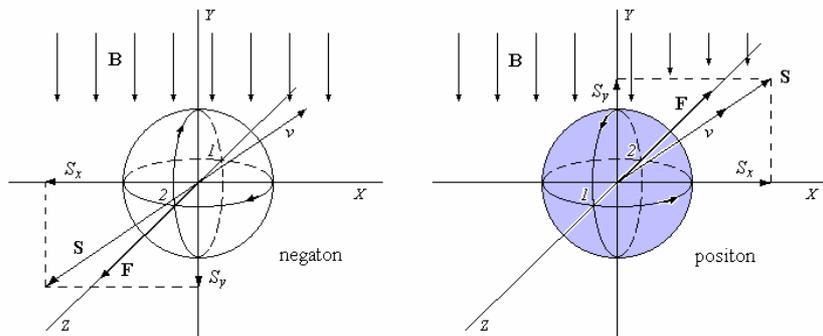

Figure 7

## 14. Conclusions

We have established a single wave-particle interaction $F_{ij}$, synthesized by expression (34), able to reproduce all the forces in Nature.

To achieve this, we started from the basis of the existence of only two kind of pulsating spherical bubbles (*positon* and *negaton*) as the ultimate constituents of matter, whose membranes oscillate according to the anharmonic potential (2). These bubbles (*tons*) produce spherical anharmonic waves (polarized field in the ether) and self-propel them in this polarized ether.

With this interaction, we obtained the electrical force for large distances and the nuclear force for very short distances.

We introduced a new concept "*Aspin*", an infinitesimal correction in the average radius of the ton consisting of an asymmetry in the size between positon and negaton, and we obtained the force of gravity as a residue of the interactions between two neutral particles composed of tons. This is true for all distances between particles that have tons with same mass. Between atoms, it is also true for all distances $d > 10^{-3}$ m but not for very short distances. For distances $d \leq 10^{-4}$ m, we obtained a stronger attraction force similar to the Casimir, van der Waals and atomic forces.

We easily modeled particles and nuclei by linking opposed tons through the interaction $F_{ij}$ and restricted values of *n*. Between positons and negatons there is always a shift of 180º, reaching maximum stability with minimum size.

Conserving the total number of tons involved in a collision, nuclear reactions can be described.

We also modeled atoms by linking nuclei with electrons in orbit through the interaction $F_{ij}$ and restricted values of *n*.

The restricted values *n* of the tons quantify matter.

Antimatter is obtained by exchanging the tons.

The pulsation of a ton contracts and dilates the ether and its rotation stretches and tenses it. The tons shape the ether. If the tons move fast, their surrounding ether will be modified continuously. We will always have a shaped ether around the ton. The configuration of the ether travels with the ton. The ton transports its wave-field. In addition, this configuration will not be spherical but will be ovoid-shaped due to the speed of the ton. From this and assumptions (a, b), we initiated electromagnetism.

Finally, let us assume that the elasticity of the ether decreases locally with the waves generated by the tons. The immediate consequence of this would be that the speed of light is less than *c* near the ton; and also that it is less among the atoms or near them. This would confirm that the reflection and the refraction of the electromagnetic waves really happen in the perturbed ether generated by the tons.

The Earth transports its own gravitational field. Away from its surface, the perturbation of the ether decreases, but on its surface the ether, though strongly perturbed, is homogeneous. The speed of light is slightly smaller than *c* but constant. As consequence, the result of the Michelson-Morley experiment is in agreement with the theory.

The Sun transports its own gravitational field. From its surface, the perturbation of the ether decreases. As consequence, light refracts near its surface.

## 15. Further Questions and Predictions

Our theory makes some predictions to be tested in the future; for example, neutral matter repels positive charges and positive charges atracts neutral matter This could be a hint towards explaining the acceleration of the expansion of the Universe, observed recently.

Due to all neutral matter in the Universe, the positively charged matter would be in the boundaries of our Universe and expanding with acceleration, at the same time dragging the unbound neutral matter (galaxies) with a smaller acceleration, and in this way there is an uniform and continuous accelerated expansion of the Universe

At the same time, neutral matter attracts negative charges and negative charge repels neutral matter. This would explain, perhaps, that negative charges lie in the surface of conductors. The negative charges find an equilibrium point in the surface of electrical conductors.

These predictions do not go against the third principle of Newton (the action-reaction principle). In fact, the description of the mechanical interaction between the ton and its surrounding ether is based on this principle. All the interactions are between tons and their surrounding ether and NOT between tons. In other words: the neutral matter forms its surrounding ether. In such ether, a ton self-propels, according the third principle of Newton. If the ton is positon (positive charge), it moves away from the neutral matter. If it is negaton (negative charge), it gets closer to the neutral matter. The neutral matter also self-propels in its ether, which is formed by tons external to her, according to the third principle of Newton. In this case, the neutral matter gets closer to positons and moves away from negatons. The final result is that, apparently, the fact following taking place: *"neutral matter repels positive charges and positive charges atracts neutral matter"* and *"neutral matter atracts negative charge and negative charge repels neutral matter"*.

Another question refers to a possible ether viscosity. In fact there is none: the tons move freely in the ether, conserving their energy. There is only trasfer of energy, from potential in the membrane to kinetic on motion and viceversa.

However, the model cannot be taken as definitive, and we suspect further refinements to be possible. The nature of them cannot be anticipated, though, at this point.

Long-standing problems of physics persist, but perhaps they are illuminated with a new light; for example, differences of masses, like electron and muon.

Finally, the ton cannot be strictly localized, because of the vibration of the membrane; consequently, a kind of uncertainty principle subsists.

## Acknowledgements

The author whises to express his gratitude to Dr. Luis J. Boya, Ana Carmona, Noemí Lana-Renault, Yann Lana-Renault, Dr. Julio Monreal, Maria Pilar Monreal and Dr. Javier Sabadell, who provided invaluable help in the development and writing of this theory.

**REFERENCES**

[1] Yoël Lana-Renault (2000). *Exact zero-energy solution for a new family of anharmonic potentials*. Rev. Academia de Ciencias. Univ. Zaragoza, Vol **55**, 103-109.
http://www.yoel-lana-renault.es/ExactzeroenergyAcadCiencias.pdf

## BIBLIOGRAPHY

Marcelo Alonso & Edward J. Finn (1987). *Fundamental University Physics*. Addison-Wesley

M. Bordag, U. Mohideen and V.M. Mostepanenko, *Phys. Reports* 353 (2001) 1 and references therein

H.B.G. Casimir. *Proc. K. Ned. Akad. Wet.* B51 (1948) 793

Raymond Chang (1991). *Chemistry*. McGraw – Hill, Inc., U.S.A.

Michel Crozon (1987). *La matière première*. Éditions du Seuil, Paris

Richard P. Feynman, Robert B. Leighton & Matthew Sands (1998). *The Feynman Lectures on Physics, Mainly Mechanics, Radiations and Heat.* Addison Wesley Longman

Richard P. Feynman, Robert B. Leighton & Matthew Sands (1998). *The Feynman Lectures on Physics, Mainly Electromagnetism and Matter, Radiations and Heat.* Addison Wesley Longman

Richard P. Feynman, Robert B. Leighton & Matthew Sands (1998). *The Feynman Lectures on Physics, Quantum Mechanics.* Addison Wesley Longman

Richard B. Firestone (1996). *Table of Isotopes.* Wiley-Interscience Publication

A. P. French (1988). *SpecialRelativity*. W.W. Norton & Company, Inc., New York

A. P. French (1993). *Vibrations and Waves*. W.W. Norton & Company, Inc., New York

Sheldon L. Glashow (1991). *The Charm of Physics*. The American Institute of Physics

Astrid Lambrecht & Serge Reynaud. *Recent Experiments on the Casimir Effect: Description and Análisis.* Preprint (2003) arXiv:quant-ph/0302073v1

S.K. Lamoreaux, Resource Letter in *Am. J. Phys.* 67 (1999) 850

G.S. Landsberg (1983). *Óptica*. Editorial MIR, Moscú

P.W. Milonni, *The quantum vacuum* (Academic, 1994)

V.M. Mostepanenko and N.N. Trunov, *The Casimir effect and its applications* (Clarendon, 1997)

Yuval Ne'eman & Yoram Kirsh (1986). *The Particle Hunters*. Sindícate of the Press of the U. of Cambridge

M.J. Sparnaay, in *Physics in the Making* eds Sarlemijn A. and Sparnaay M.J. (North-Holland, 1989) 235 and references therein

Hubert Reeves (1988*). Patience dans l'Azur. L'évolution cosmique*. Éditions du Seuil, Paris

Robert Resnick (1981). *Basic Concepts in relativity and early quantum Theory*. John Wiley & Sons, Inc.

Robert Resnick (1981). *Introduction to Special Relativity*. John Wiley & Sons, Inc.

Leonard I. Schiff (1968). *Quantum Mechanics*. McGraw-Hill